\documentclass[fleqn,usenatbib]{rasti}
\usepackage{newtxtext,newtxmath}
\usepackage[T1]{fontenc}
\DeclareRobustCommand{\VAN}[3]{#2}
\let\VANthebibliography\thebibliography
\def\thebibliography{\DeclareRobustCommand{\VAN}[3]{##3}\VANthebibliography}
\usepackage{graphicx}	
\usepackage{amsmath}

\usepackage{orcidlink}

\title[Policy Based Radiative Transfer]{Solving the $2$-Level Atom Non-LTE Problem using Soft Actor-Critic Reinforcement Learning}

\author[B. Panos and I. Mili\'{c}]{
  Brandon Panos\orcidlink{0000-0002-7096-7941}$^{1}$\thanks{E-mail: \href{mailto:brandon.panos@fhnw.ch}{brandon.panos@fhnw.ch}},\ 
  Ivan Mili\'{c}\orcidlink{0000-0002-0189-5550}$^{2}$\\\
  $^{1}$University of Applied Sciences and Arts Northwestern Switzerland (FHNW), Bahnhofstrasse 6, 5210 Windisch, Switzerland\\
  $^{2}$Institute for Solar Physics (KIS), Georges-K\"{o}hler-Allee 401a, 79110 Freiburg, Germany}

\date{Accepted XXX. Received YYY; in original form ZZZ}
\pubyear{2025}

\begin{document}
\label{firstpage}
\pagerange{\pageref{firstpage}--\pageref{lastpage}}
\maketitle

\begin{abstract}
We present a novel reinforcement learning (RL) approach for solving the classical 2-level atom non-LTE radiative transfer problem by framing it as a control task in which an RL agent learns a depth-dependent source function $S(\tau)$ that self-consistently satisfies the equation of statistical equilibrium (SE). The agent's policy is optimized entirely via reward-based interactions with a radiative transfer engine, without explicit knowledge of the ground truth. This method bypasses the need for constructing approximate lambda operators ($\Lambda^*$) common in accelerated iterative schemes. Additionally, it requires no extensive precomputed labeled datasets to extract a supervisory signal, and avoids backpropagating gradients through the complex RT solver itself. Finally, we show through experiment that a simple feedforward neural network trained greedily cannot solve for SE, possibly due to the moving target nature of the problem. Our $\Lambda^*-\text{Free}$ method offers potential advantages for complex scenarios (e.g., atmospheres with enhanced velocity fields, multi-dimensional geometries, or complex microphysics) where $\Lambda^*$ construction or solver differentiability is challenging. Additionally, the agent can be incentivized to find more efficient policies by manipulating the discount factor, leading to a reprioritization of immediate rewards. If demonstrated to generalize past its training data, this RL framework could serve as an alternative or accelerated formalism to achieve SE. To the best of our knowledge, this study represents the first application of reinforcement learning in solar physics that directly solves for a fundamental physical constraint. 
\end{abstract}


\section{Introduction}

Radiative transfer (RT) calculations are critical for interpreting light from diverse astronomical sources, including the early universe \citep{Michele2023}, supernovae \citep{supernova2022}, galaxy formations \citep{rt_galaxy_2023}, stars, and our Sun \citep{RH2001}. Researchers infer physical variables (e.g., magnetic field strength, temperature, density, velocity) by analyzing spectra from specific atomic transitions sensitive to these local conditions. The inference process typically involves iteratively running forward RT simulations and adjusting parameters until the calculated spectrum matches the observation. The resulting parameters are then hypothesized to represent the object's physical state. This method crucially assumes that the simulation accurately captures all necessary physics. However, rigorously modelling the radiation-matter interaction, especially under non-Local Thermodynamic Equilibrium (non-LTE) conditions relevant for many spectral lines, is often computationally intractable. The non-LTE conditions denote any departure of the system from the state of LTE and directly imply a strong, nonlinear coupling between the state of gas and radiation. Understanding this coupling is also extremely important in the context of radiative (magneto) hydrodynamical simulations, where the radiation also serves as a crucial means of energy transport \citep[e.g.][]{MuramChromo, BIFROST}. As a result, researchers frequently adopt unwarranted simplifying assumptions such as plane-parallel geometry (effective $1$D atmospheres), complete frequency redistribution (CRD), stationary backgrounds, low-order numerical methods, and coarse spatial grids, which all call into question the reliability of the inferred physical parameters.  

The problem of RT involves solving a massive coupled set of linear equations, which in the non-LTE regime is done iteratively using techniques such as the Lambda Iteration (LI) or Accelerated Lambda Iteration (ALI) \citep{ali1, ali2, ali3}, Gauss-Seidel \citep{gs_method1995}, implicit lambda \citep{Implicit_Lambda2014}, multi-grid \citep{Grid1_1991, Grid2_2013}, bi-conjugate gradient methods \citep{Bi_conjugate_Gradient_2009}, and matrix-free approaches \citep{matrix_free2023}. Despite algorithmic progress, accurate and complete simulations still remain computationally prohibitive. 

For this reason, researchers have started using machine learning-based methods to accelerate RT simulations, including learning a mapping between LTE and non-LTE level populations in $3$D \citep{sunny_net2022}, predicting the departure coefficients of atomic level populations \citep{Graph2022}, and replacing the computational core of simulations with physics-informed neural networks (PINNs). Direct emulation of the simulation engine has resulted in impressive offline performance gains for numerical weather predictions \citep{PINN2021_weather}, the forward modeling of galaxy spectral energy distributions \citep{rt_galaxy_2023}, and accurate predictions of the complete $4$D hydrogen fraction evolution of the Epoch of Reionization \citep{Michele2023}. Additionally, neural fields and PINNs have been applied to the inverse problem in solar physics, specifically for spectropolarimetric inversions to infer atmospheric parameters like the magnetic field \citep{carlos_neural_fields_inversions_2025, momo_pinns_2025}. Despite these advancements, PINNs are still brittle under generalization, do not scale well, and struggle to function as RT emulators in real-time online tasks \citep{torch_adaptor2023}.\\ 

Reinforcement learning (RL) represents a powerful branch of machine learning that remains underutilized within the sciences despite possessing qualities seemingly highly advantageous for simulation tasks. Many computational physics problems involve iterative schemes to reach a solution, which, when employing supervised function approximation or emulation, can result in a fast accumulation of errors. Unlike most approximation paradigms, RL optimizes policies for sequential decision-making by maximizing a cumulative long-term reward signal. This framework inherently balances exploration (sampling novel strategies) and exploitation (leveraging effective known strategies), enabling the potential discovery of non-intuitive or more efficient solutions. Additionally, the design of the reward function provides flexibility that can promote the exploration of fundamentally new solutions in the case of a reward sparse framework (usually requiring more episodes and therefore compute), while dense, structured rewards can incorporate domain knowledge which forces the policy into the realm of imitation or expert learning \citep{imitation_leanring2024}. RL has achieved unprecedented success within the gaming industry thanks to its ability to automate the algorithmic discovery process, resulting in superhuman performance in complex strategic games such as Chess \citep{AlphaZero2017}, Go \citep{GO2016}, and Dota 2 \citep{Dota2019}. By formulating scientific problems within an RL framework ("gamification"),  researchers have achieved substantial breakthroughs, including predicting protein structures with atomic-level accuracy \citep{alphafold2021}, developing novel turbulence modeling strategies via multi-agent RL \citep{turbulence2020}, and discovering faster algorithms for fundamental basic computations such as matrix multiplication \citep{AlphaTensor2022}. These examples illustrate the potential of RL in the sciences, beyond traditional game environments. 

In this work, we propose a novel solution to the RT problem that dispenses with manual heuristics and avoids constructing labeled datasets for supervision. Instead, our approach relies exclusively on a carefully designed reward function that allows an RL agent to develop a policy that can efficiently drive a simple non‐LTE atmosphere into equilibrium, purely through reward‐based interaction with a physics engine. Whether this framework generalizes to other atmospheric profiles and more realistic scenarios remains a task for future studies.

\section{Two-level atom non-LTE problem}

For a proof of concept, we selected a simple academic, non-LTE problem: a single species of $2$-level atom under the assumption of complete frequency redistribution (CRD) in a $1$D plane-parallel atmosphere. The depth grid is defined on $\log_{10}(\tau_c)$ ranging from $-7$ to $2$ consisting of $91$ depth points, where $\tau_c$ is the continuum optical depth. For simplicity, the atmosphere is assumed to be isothermal, represented by a constant Planck function, $B(\tau_c) = B = 1$. A Gaussian (Doppler) profile $\phi(\nu)$ is used to model the absorption line and is taken to be constant with depth. Frequency and angle integrations employ Gaussian quadrature schemes. 

The radiative transfer equation (RTE) in this case reads: 
\begin{equation}
    \mu \frac{dI(\tau_c, \mu, \nu)}{d\tau_c} = \left[1+\eta_0\phi(\nu)\right] \left[I(\tau_c,\mu,\nu) - S(\tau_c)\right],
\end{equation}
where $\eta_0$ denotes the ratio of line-integrated opacity to continuum opacity and was set to $\eta_0 = 10^3$. RTE was solved numerically along discrete rays ($\mu=$const) using a second-order short characteristics method \citep{OLSON1987325} to compute the specific intensity $I(\tau_c, \mu, \nu)$.  Finally, to mimic a solar atmosphere, we set the boundary conditions such that the incoming radiation ($\mu <0$) at the top of the atmosphere was given by $I_\text{in}=0$ and the outgoing radiation ($\mu>0$) at the bottom of the atmosphere was provided by the Planck function $I_\text{out}=B$. The core of the non-LTE problem lies in satisfying the SE equation, which for a $2$-level atom is given by:
\begin{equation}
S(\tau_c) = (1-\epsilon) \bar{J}(\tau_c) + \epsilon B(\tau_c)
\label{se_eqn}
\end{equation}
where $\bar{J}$ is the so-called scattering integral, i.e., absorption profile-averaged mean intensity:
\begin{equation}
\bar{J}(\tau_c) = \frac{1}{2} \int \phi(\nu)  \int_{-1}^{1} I(\tau_c, \mu, \nu) d\mu ~ d\nu.
\label{rad_field_eqn}
\end{equation}
The value of the photon destruction probability $\epsilon$ in Eq.\ref{se_eqn} dictates whether we are in a high-scattering or strongly absorbing region of the atmosphere. When $0 \leq \epsilon \ll 1$, the source function decouples from the Planck function and is determined by the radiation field and non-local contributions. However, when $\epsilon \to 1$, the source function is entirely locally determined, i.e, we recover the assumption of LTE. We set $\epsilon = 10^{-4}$ to be constant with depth. The setup is similar to very well-known problems used to benchmark the operator perturbation techniques \citep[see e.g.][]{OLSON1987325, Mihalasbook1978}. Note that the problem is linear in the Source function and, after suitable discretization, can be solved directly. However, it serves as a pedagogical introduction to multi-level problems which can be highly non-linear \citep[][]{Mihalasbook1978, ali1}. 

A classical, simple solution of these problems is to iterate equations \ref{rad_field_eqn} and \ref{se_eqn} until convergence. This is known as ``Lambda iteration''. For small values of $\epsilon$, the procedure is extremely slow and converges linearly, which makes defining a stopping criterion difficult \citep[a thorough discussion is presented in the textbook of ][]{Mihalasbook1978}. A widely accepted approach is to accelerate this procedure by the operator perturbation techniques, which are, in a way, equivalent to preconditioning \citep[e.g.][]{Mihalasbook1978, OLSON1987325, Bi_conjugate_Gradient_2009}. The core of all these methods is the process of solving RTE for all necessary directions and frequencies, and then updating the source function (Eq.\,\ref{se_eqn}). Regardless of the method employed, we refer to this process as one $\Lambda$ iteration.

\section{Reinforcement Learning Formulation}

\begin{figure}
\centering
\includegraphics[width=0.48\textwidth]{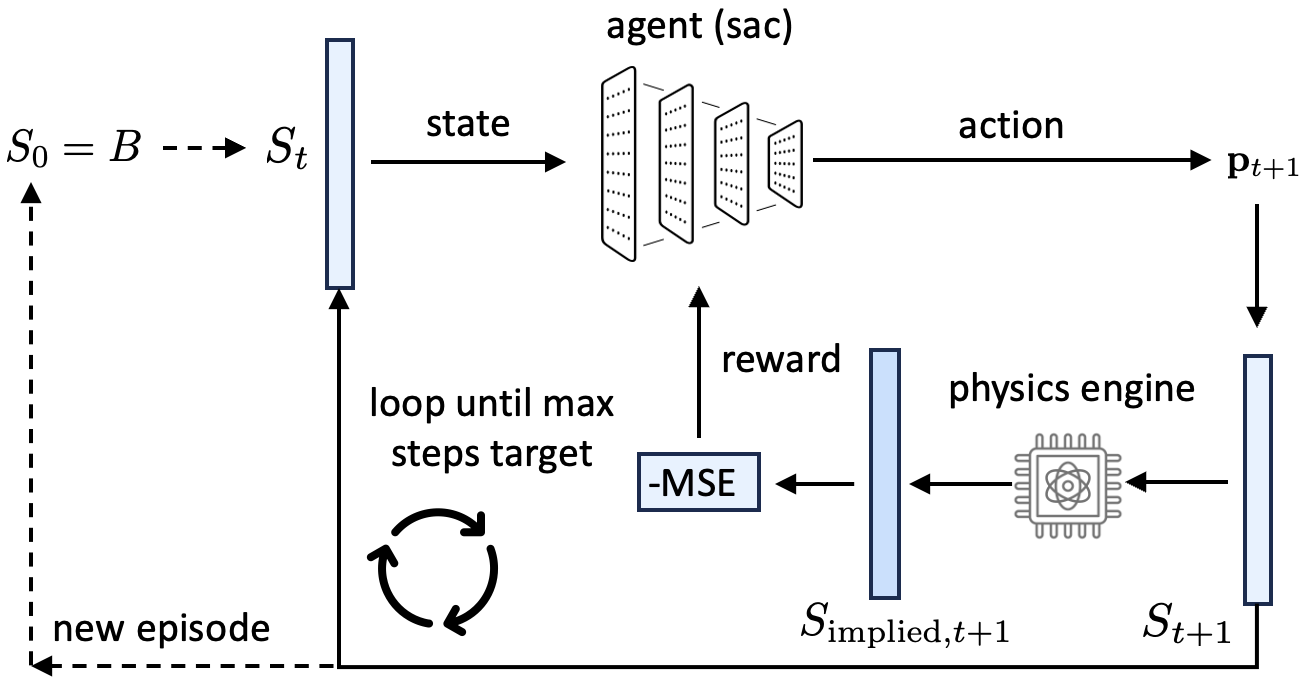}
\caption{Diagram of the training loop: In a clockwise fashion; the source function is initiated to the Planck function $B$, which is then sent to the agent as a state. The agent's policy decides on an action that generates four parameters $\mathbf{p}$ that are used to construct a smooth, well-behaved source function across the entire depth scale. The agent's predicted solution is passed to the physics engine, which generates a conditioned "implied" source function that would satisfy SE. The residual of the agent and implied source is used as a reward signal to instruct the agent's policy. The inner loop iterates until either a max step criterion is reached or the agent obtains the target, defining a single episode. Once the inner loop terminates, the source function is once again initiated to the Planck function, and the agent can try and refine its policy.}
\label{rl_loop_image}
\end{figure}

We reformulate the RT problem within an RL framework, which models the interaction between an agent and an environment. In this framework, the agent selects actions to influence the environment's state transitions, aiming to learn a policy $\pi(a|s)$ that maximizes cumulative future rewards. For the RT problem, the environment represents the physical system, with its state characterized by the current depth-dependent source function $S(\tau_c)$. The agent is implemented using the Soft Actor-Critic (SAC) algorithm. Unlike traditional methods that modify $S(\tau_c)$ at discrete points, our agent's action is the selection of a low-dimensional parameter vector $\mathbf{a} \in [-1, 1]^4$. This vector is scaled to define four physical parameters, $\mathbf{p} = (\text{floor, amplitude, center, width})$, which collectively determine the source function across the entire depth grid via a smooth, generally monotonic sigmoid profile:
\begin{equation}
S(\tau_c; \mathbf{p}) = \text{floor} + \text{amplitude} \times \text{sigmoid}\left( \frac{\log_{10}(\tau_c) - \text{center}}{\text{width}} \right).
\label{source_param_eqn}
\end{equation}
The agent, as seen in Fig. \ref{rl_loop_image}, observes the resulting source function $S(\tau_c; \mathbf{p})$ generated by its chosen parameters. The core learning mechanism is the reward function, designed to drive the agent towards the SE solution. At each step $t$, given the agent's proposed parameters $\mathbf{p}_t$ yielding $S_t = S(\tau_c; \mathbf{p}_t)$, the environment calculates the corresponding mean intensity $\bar{J}(S_t)$ and determines the source function implied by SE:
\begin{equation}
S_{\text{implied}, t} = (1-\epsilon) \bar{J}(S_t) + \epsilon B.
\end{equation}
One such step is equivalent to one $\Lambda$ iteration in conventional approaches. The reward $R_t$ is the negative mean squared error (MSE) between the parameterized and implied source functions:
\begin{equation}
R_t =  - \frac{1}{N_D} \sum_{i=1}^{N_D} [S_t(\tau_{c,i}) - S_{\text{implied}, t}(\tau_{c,i})]^2.
\label{reward_eqn}
\end{equation}
Maximizing this reward incentivizes the agent to find parameters $\mathbf{p}$ such that $S(\tau_c; \mathbf{p})$ globally satisfies the SE condition. An episode involves the agent iteratively proposing parameters and receiving rewards, terminating either after a maximum number of iterations or when the MSE between the agent's current $S_t$ and a pre-computed, converged ALI solution falls below a threshold. The ALI solution serves solely as a termination criterion and does not inform the reward signal during training. Note that all rewards are $\leq 0$, meaning the agent tries to obtain a net reward that approaches zero over the maximum 50 permitted time steps of exploration (inner loop counter). The problem of finding SE is what is termed a moving target problem within the domain of RL \citep{moving_target2015human}, because the implied $S$ used to calculate the residual depends on the actors actions. Moving target problems are notoriously difficult for direct gradient methods to solve, because the loss landscape continuously reshapes itself around the latest guess, resulting in noisy gradients \citep{moving_target2015}. 

\begin{figure}
\centering
\includegraphics[width=0.48\textwidth]{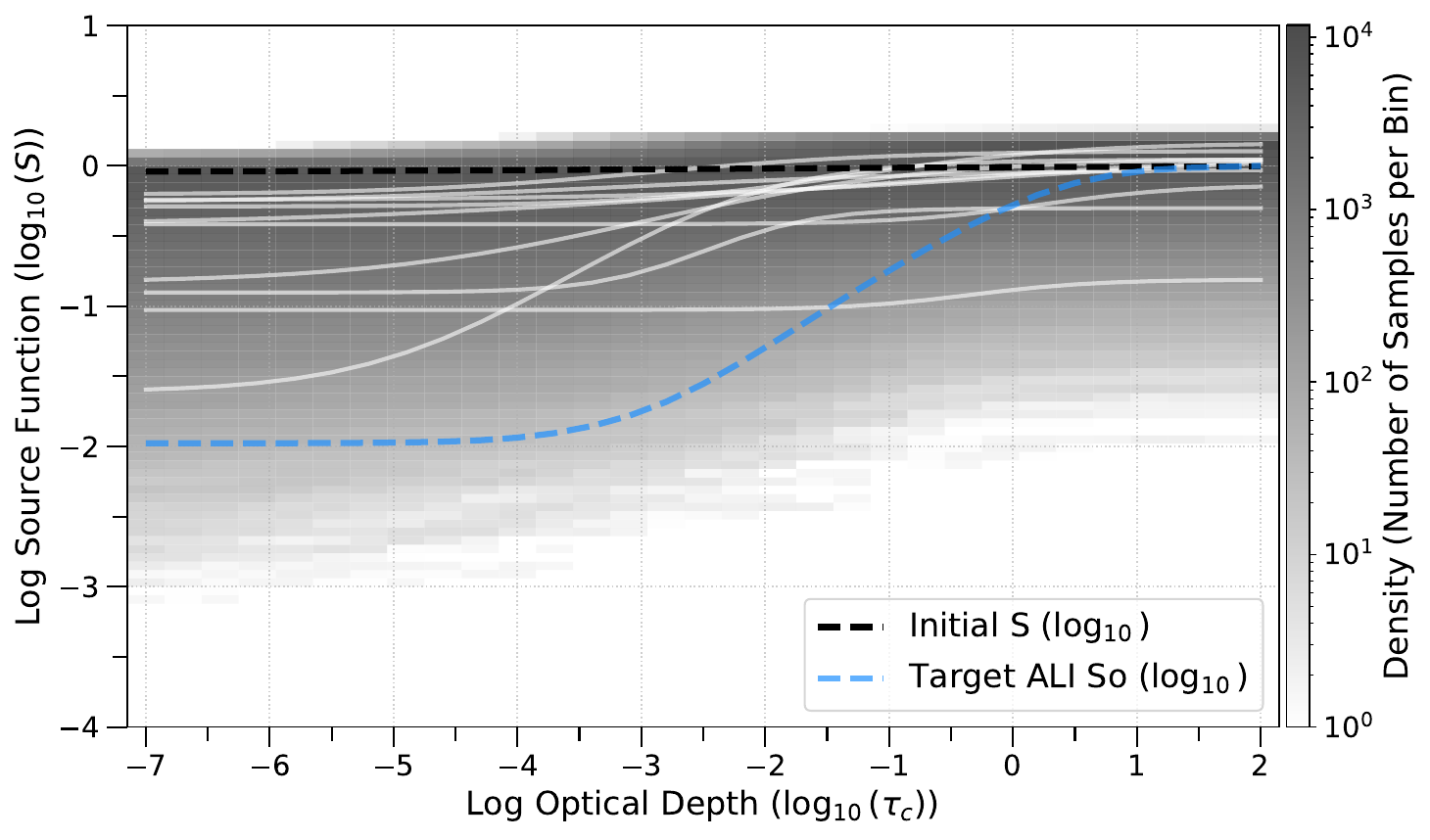}
\caption{Reachable solution space of the parameterized source function. The density heatmap (log scale) shows the frequency of $\log_{10}(S)$ values versus $\log_{10}(\tau_c)$ based on $50,000$ random parameter samples. Overlays show the initial guess (black), target ALI solution (blue dashed), and example random profiles (light gray).}
\label{solution_space_image}
\end{figure}

Figure \ref{solution_space_image} illustrates the flexibility of the parameterized source function by visualizing its solution space. The background heatmap represents the probability density, computed from $50,000$ uniformly sampled parameter sets and displayed on a logarithmic scale. This reachable space is compared against the initial guess (dashed black line) and the desired target solution obtained from ALI (dashed blue line). Ten randomly selected examples of the parameterized function (gray solid lines) illustrate some of the characteristic shapes generated within the defined parameter boundaries, showing the parameterization's ability (or limitations) in approximating the target profile across different optical depths. Due to its limited flexibility, there will inevitably be a small residual between the target and the best solution, as verified by a Levenberg-Marquardt least squares fit. This implies that our agent is exploring the solution space to effectively find the infimum of the residual.

\section{The Soft Actor-Critic Algorithm}
The agent depicted in Fig. \ref{rl_loop_image} represents a state-of-the-art, off-policy, stochastic, model-free RL algorithm called Soft Actor-Critic (SAC), which uniquely addresses problems with continuous action spaces \citep{sac_2018}. It is composed of an ensemble of backpropagation function approximators working in tandem to extract an efficient policy. SAC is based on the maximum entropy actor-critic framework, which is both sample efficient (due to an experience replay buffer) and insensitive to hyperparameters. The actor maps the observed state (the $S$-profile) to a probability distribution over actions (the parameters $\mathbf{p}$), while the critic (Value Function) estimates the expected future cumulative reward (Q-value) for taking a given action in a given state. In practice, two main and two target critics are used to stabilize training. SAC optimizes a policy $\pi$ to maximize a trade-off between the expected cumulative reward and the policy's entropy $H$:  

\begin{equation}
J(\pi) = \sum_{t=0}^{T} E_{(\mathbf{s}_t, \mathbf{a}_t) \sim \rho_\pi} \left[ R(\mathbf{s}_t, \mathbf{a}_t) + \alpha H(\pi(\cdot|\mathbf{s}_t)) \right],
\label{sac_objective}
\end{equation}

where $\mathbf{s}_t$ is the state (S-profile), $\mathbf{a}_t$ is the action (parameters), $\rho_\pi$ is the state-action distribution induced by policy $\pi$, $R$ is the reward from Eq. \ref{reward_eqn}, and $\alpha$ is a temperature parameter balancing reward maximization and entropy maximization. Effectively, Eq. \ref{sac_objective} instructs the agent to find the optimal policy while behaving as randomly as possible. If the entropy is maximized then each of the four variables will be sampled from a flat distribution, implying zero strategic decision making and consequently a low net reward. On the other hand, maximizing the reward directly could result in suboptimal behaviour due to insufficient exploration. Finding a balance between these two tensions (which the algorithm regulates internally) encourages a healthy exploration and prevents the policy from collapsing to a deterministic, potentially suboptimal solution, which as we will see, might be an important attribute that allows SAC to solve this particular moving target problem. 

\begin{figure}
\centering
\includegraphics[width=0.49\textwidth]{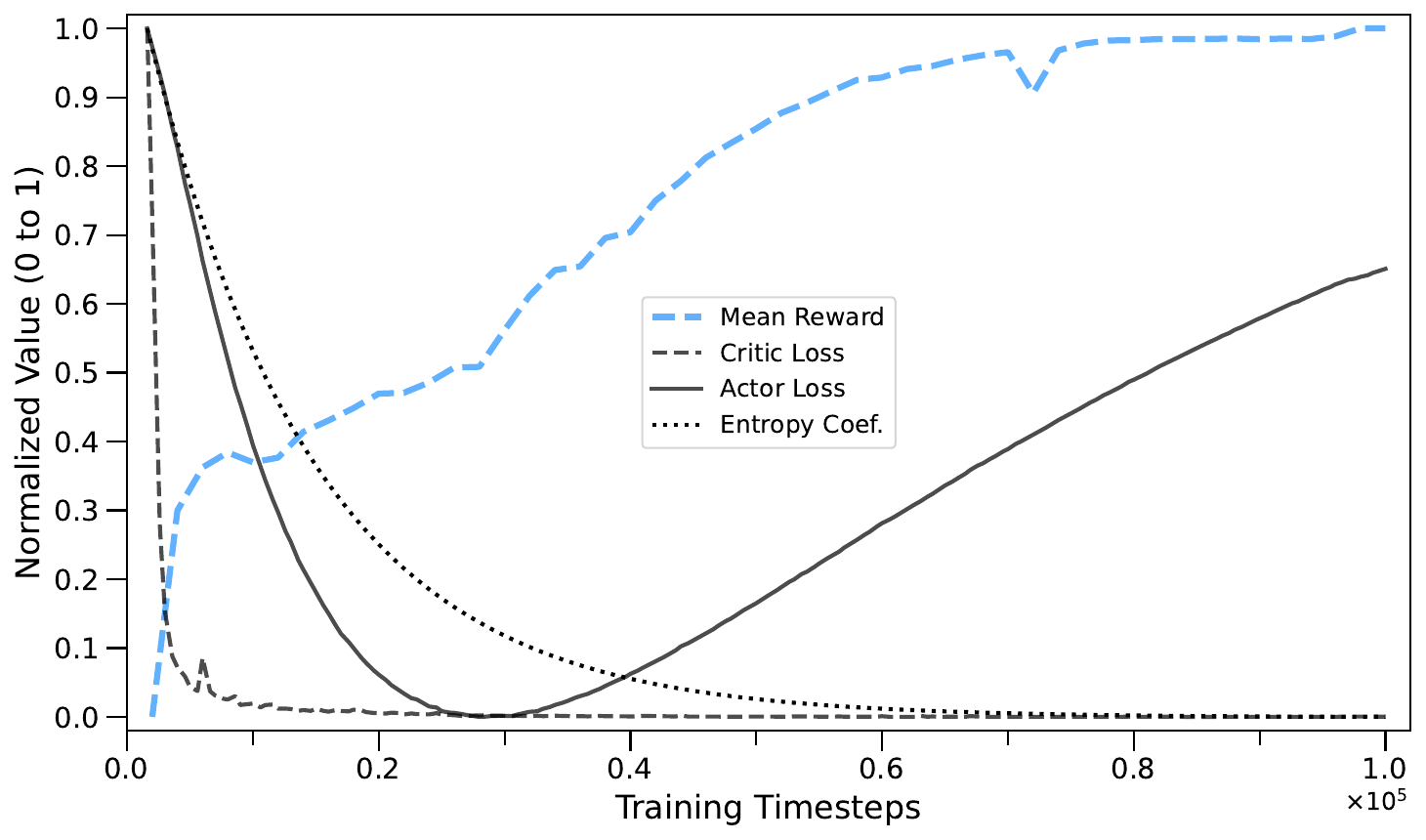}
\caption{Training performance of the SAC agent. The figure shows the mean normalized reward, critic loss, actor loss, and entropy coefficient $\alpha$, as a function of training steps. The steadily increasing reward demonstrates successful learning, while the critic loss decreases, indicating convergence of the value function estimate. The decreasing entropy coefficient signifies a shift from exploration towards policy exploitation.}
\label{learning_curves_image}
\end{figure}

The SAC agent was trained using Stable Baselines3 \citep{stable_baselines} for a specified number of total timesteps. The training progress was monitored, and the best-performing model (based on evaluation episodes) was saved. The learning dynamics of the actor-critic system can be seen in Fig. \ref{learning_curves_image}. Here, the steadily increasing mean reward (blue dashed line) indicates successful learning, as the SAC algorithm extracts an efficient policy that finds a set of parameters $\{\mathbf{p}\}$ which populate and cluster around the target SE solution, resulting in high relative accumulative rewards. The critic networks (two main and two target) learn to approximate the Q-function. The decrease in loss signifies that the critic's predictions of future cumulative rewards are becoming more accurate based on the observed transitions and rewards. The solid black line represents the loss associated with the actor network (the policy), which seeks to output actions that maximize the expected Q-value estimated by the critic. Its initial decrease corresponds to the policy rapidly improving, while the subsequent increase is not uncommon and does not necessarily indicate poor performance, as evidenced by the steady increase in mean reward. It is possible for the policy loss to both increase while improving the estimate of the source function. This effect is consistent with the mechanics of the SAC algorithm \citep{sac_2018,sac_applications_2018}, and simply reflects the changing scale of the actor's objective function as the improved policy guides the trajectory through increasingly rewarding regions of the state space. Finally, the black dotted line shows the entropy coefficient, $\alpha$, which weights the importance of the policy's entropy in the objective function. The trend shows that the agent starts with an aggressive exploration (high entropy) strategy, and then over the course of training, slowly focuses more on exploiting the known good actions (lower entropy). 

\section{Results - Comparing SAC Policy Performance}

\begin{figure}
\centering
\begin{minipage}{0.49\textwidth}
  \centering
  \includegraphics[width=\textwidth]{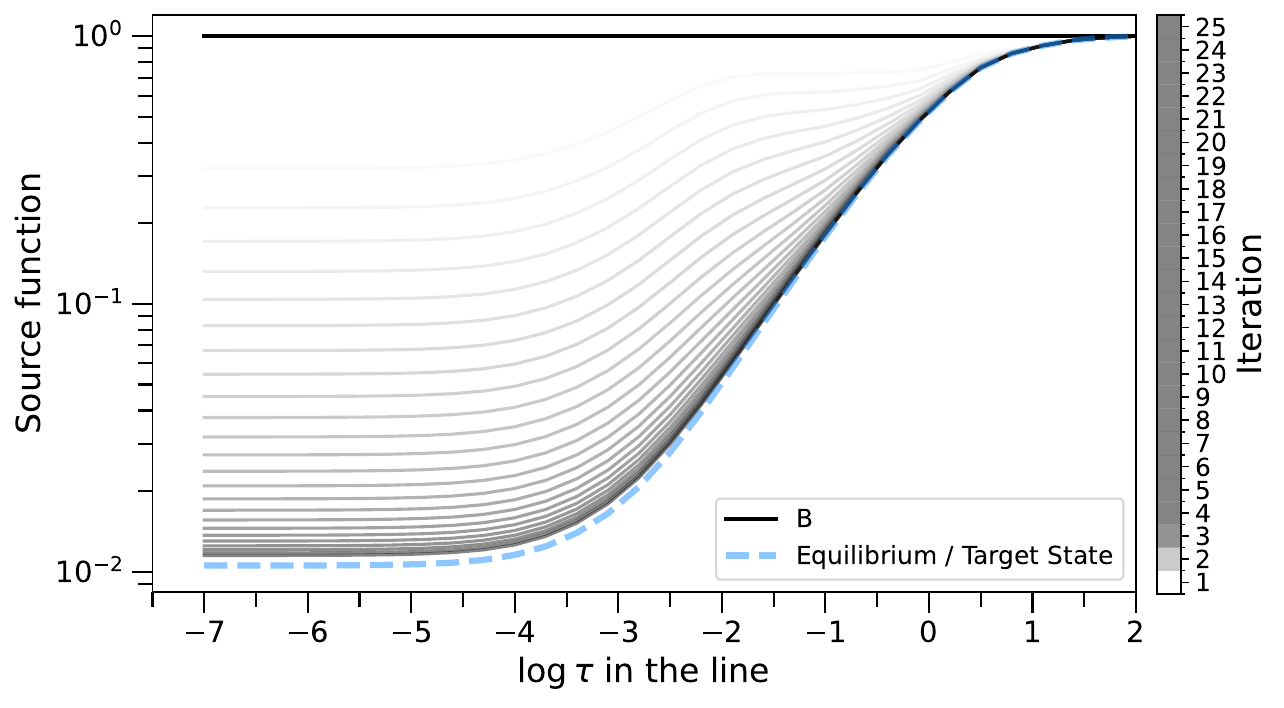}
\end{minipage}
\hfill
\begin{minipage}{0.49\textwidth}
  \centering
  \includegraphics[width=\textwidth]{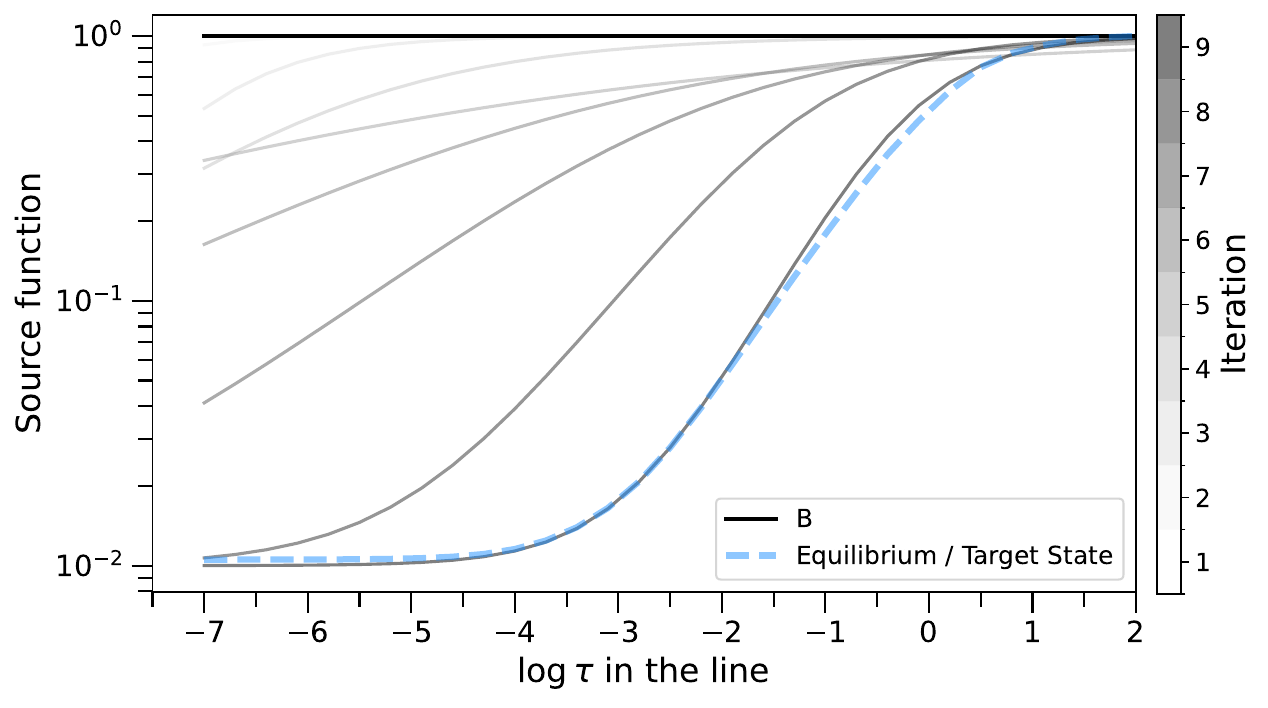}
\end{minipage}
\hfill
\begin{minipage}{0.49\textwidth}
  \centering
  \includegraphics[width=\textwidth]{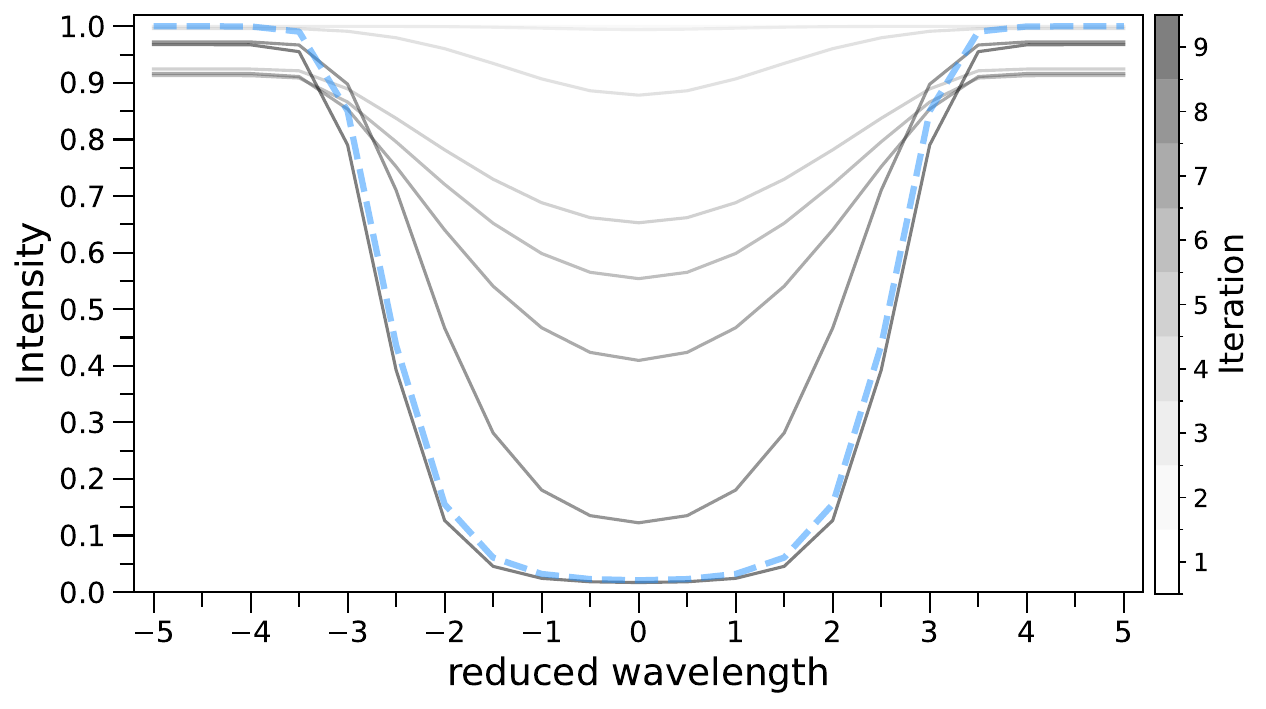}
\end{minipage}
\caption{Upper panel: ALI method converges to SE (dashed blue line) after multiple iterations. Middle panel: The SAC policy drives the simple non-LTE simulation into SE with fewer iterations. Decreasing the discount factor promotes policies that converge faster. Lower Panel: Evolution of line-of-sight ($\mu=1$) observed intensity as a function of the agents policy.}
\label{comparison_image}
\end{figure}

We evaluate the performance of the trained SAC agent by comparing its convergence behavior to the ALI scheme. The upper panel of Fig. \ref{comparison_image} depicts the standard ALI convergence, starting from an initial guess where the source function equals the Planck function ($S=B$, solid black line). ALI iteratively refines the source function profile across the optical depth scale. The grayscale lines represent successive iterations, gradually approaching the target SE state (dashed blue line). 

The middle panel depicts the same process, but in this case, the source function is driven by the trained SAC agent's optimal policy, corresponding to a single evaluation episode within the inner loop of Fig. \ref{rl_loop_image}. The agent begins with an initial state ($S=B$) and, at each step, selects an action (a set of four parameters $\mathbf{p}$) based on its learned policy $\pi^*(a|s)$. This action generates a parameterized source function according to Eq. \ref{source_param_eqn}, represented by the grayscale lines. The plot demonstrates that the SAC policy rapidly drives the system towards the target SE solution (dashed blue line), achieving the same tolerance as the ALI scheme with fewer steps, representing a significant reduction in the number of solver interactions.

Finally, the lower panel shows the corresponding evolution of the emergent line-of-sight intensity profile, $I(\mu=1)$, as a function of wavelength (normalized to Doppler width). As the SAC policy adjusts the source function parameters step-by-step (middle panel), the resulting spectrum evolves from the initial flat continuum ($I=B=1$) towards the final absorption line expected at convergence (dashed blue line), providing a visual confirmation of the method's legitimacy. 

Furthermore, the agent can be incentivized to find more optimal policies by reducing the discount factor $\gamma < 0.99$, which serves as a prefactor for the trajectory's reward chain $\gamma^t r_t$, leading to a prioritization of immediate returns. It is unclear how the choice of discount factor will impact the agent's ability to generalize and find optimal solutions. We hypothesize that the agent's optimal solution would be compromised with lower $\gamma$ as exploration will be dampened. 

It is important to note that this comparison focuses on the number of iterations required after training. The RL agent requires a separate, potentially computationally intensive, training phase. The performance only has meaningful acceleration potential if the given policy generalizes to other unseen atmospheric configurations.

\subsection{Policy Stability and Comparison with Direct Optimization}

\begin{figure}
\centering
\includegraphics[width=0.49\textwidth]{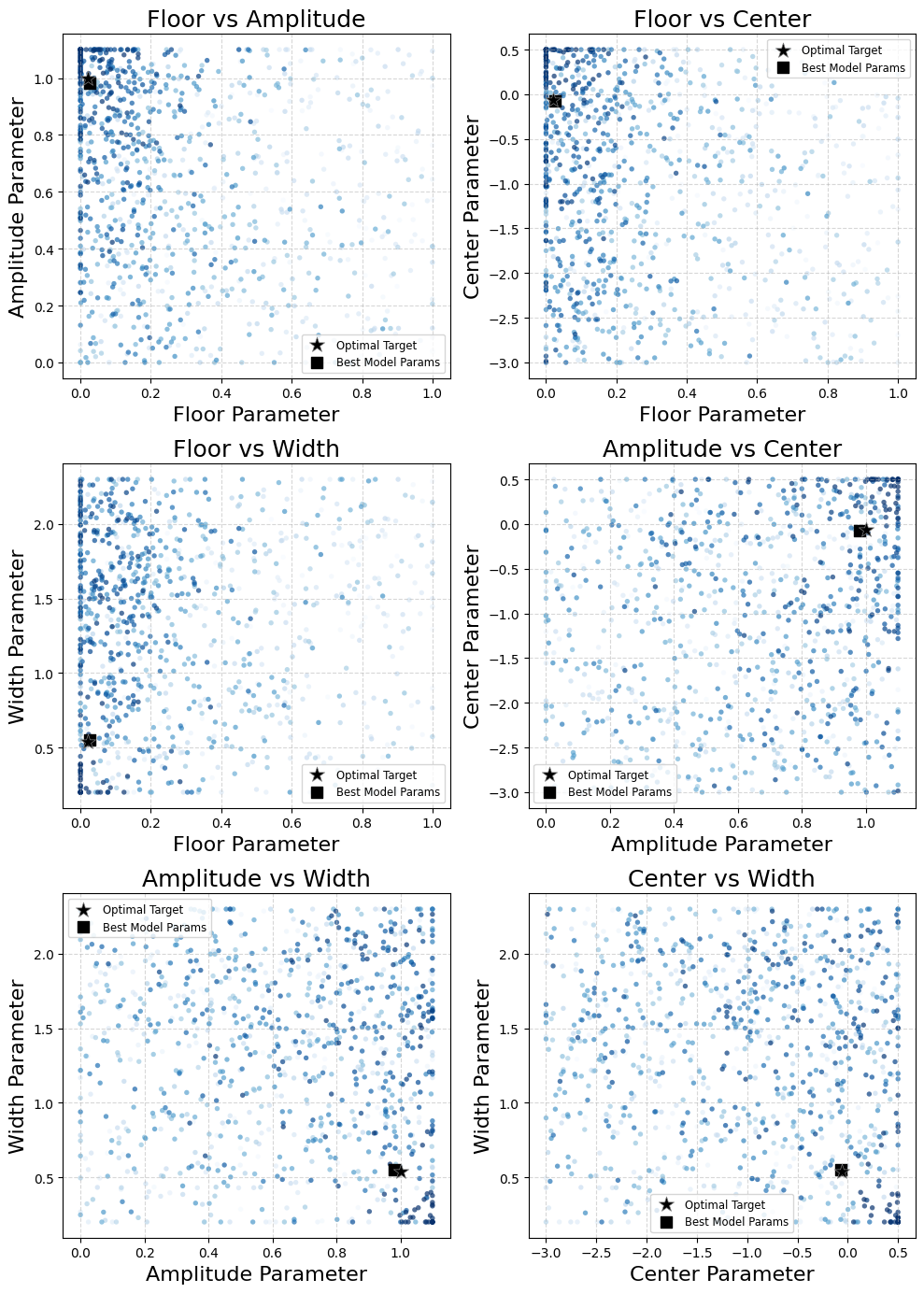}
\caption{Evolution of parameters during SAC training: The six phase planes show how the source function parameters change over time, with darker blue dots indicating later agent actions. The target solution of SE in the parameter space is indicated by a black star, while the policy's optimal solution is indicated by a black square. The plot shows how the parameters migrate during training towards the target setting.}
\label{sac_parameters_image}
\end{figure}

\begin{figure}
\centering
\includegraphics[width=0.49\textwidth]{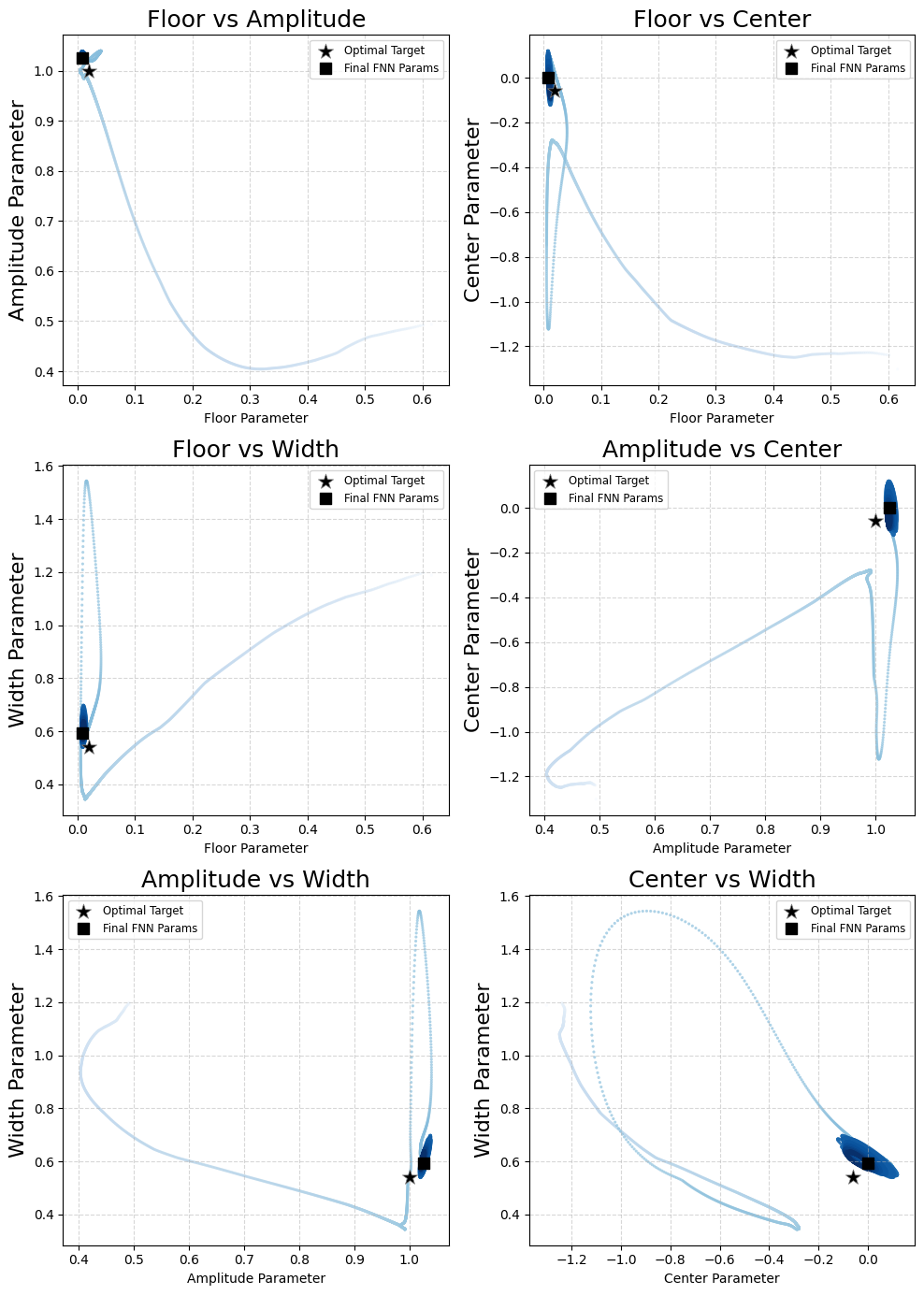}
\caption{Evolution of parameters using direct optimization: Same as Fig. \ref{sac_parameters_image}. In this case, the simple FNN directly seeks the solution; however, due to the moving target nature of the problem, the model gets trapped in a suboptimal region of the parameter space, a behaviour that is consistent regardless of model complexity or training time.} 
\label{direct_parameters_image}
\end{figure}

Extracting RL policies for an optimization task often requires a massive number of agent-environment interactions and is therefore computationally expensive. It then becomes important to demonstrate superior performance over simpler, more direct methods. To test the benefits of a sophisticated, long time horizon, delayed-reward planning algorithm over an essentially greedy policy, we replaced the SAC agent in Fig, \ref{rl_loop_image} with a fully connected feed forward network (FNN), while keeping the remaining logic of the loop consistent. The FNN was trained using the same residual-based feedback loop. The performance and stability of these two approaches differ significantly, as illustrated by comparing the evolution of the source function parameters during training.

Figure \ref{sac_parameters_image} visualizes the trajectory of the four source function parameters $(\text{floor, amplitude, center, width})$ explored by the SAC agent during training, displayed across the six pairwise parameter phase planes. Each blue dot represents the parameter set generated by an agent action, with the color darkening over time to indicate later steps in the training process. 

The agent initially explores a wide region of the parameter space (lighter dots) before progressively clustering around the target SE solution represented by the black star in each panel. This behavioural tendency from exploration into exploitation is a signature characteristic of SAC and is consistent with the decreasing entropy curve in Fig. \ref{learning_curves_image}. The plot clearly shows the agent effectively navigating the parameter space, with the parameter samples migrating and clustering around the policy's optimum (black squares), which closely approximates the true target. The minor difference between the policy's optimum and target is likely due to the aforementioned limitations of the sigmoid function, since its lack of flexibility results in at best the discovery of an infimum. This fixed best case error is also likely responsible for pushing the model's parameter search to its boundaries, resulting in a darkening of the plot edges, as SAC searches for a better configuration that does not exist.

In contrast, Fig. \ref{direct_parameters_image} illustrates the parameter evolution when using the direct FNN optimization approach. Unlike the stochastic nature of SAC, the FNN greedy optimization scheme performed a controlled, rapid, and direct parameter trajectory toward the optimal settings, in far fewer learning cycles, however, the model never obtains the target, but oscillates in a neighbourhood close to the true solution. Note that the optimal setting is not degenerate, and that all six parameters have to therefore be aligned with the stars for an optimal solution.

We hypothesise that the moving target nature of the problem, where the target for the model depends on its output, results in a continuous undulation and restructuring of the loss landscape, something that proves difficult for standard backpropagation FNN to learn, regardless of the model's complexity or training time.

 SAC on the other hand decouples policy improvement from value estimation, where the critic bootstraps value estimates via the Bellman equation and smooths them with soft targets using Polyak averaging, generating a low‐variance learning signal, even within a moving target scenario. The actor can then separately optimize against these stable targets. Additionally, the entropy-driven search of SAC allows it to escape local optima that trap the FNN. 

\section{Conclusion and Outlook}

In this work, we have presented a novel reinforcement learning solution using SAC for the classical 2-level atom non-LTE radiative transfer problem. By framing SE as a control task, an agent learns an optimal policy to parameterize the source function $S(\tau_c)$ based solely on reward signals derived from the SE residual, obtained via interaction with a formal solver. This method does not require direct knowledge of the ground truth, nor does it require a labeled dataset or the backpropagation of gradients through the RT solver itself. Our results demonstrate that the agent successfully learns a policy satisfying the SE constraint for a simplified solar atmospheric model. Once trained, this policy drives the system to equilibrium in significantly fewer iterations than a standard ALI scheme for the tested configuration, suggesting a potential for computational acceleration. Additionally, the SAC method can handle the moving target nature of the iterative SE problem, converging reliably where a direct feedforward network optimization failed due to instabilities in the gradient signal. To the best of our knowledge, this represents the first application of RL to directly enforce SE in solar physics or radiative transfer. Our method offers an "Approximate-Lambda-Operator-Free" path to equilibrium, while the formal solve to obtain the mean intensity $\bar{J}$ (the result of the true Lambda operator) is necessary to structure the reward function, the technique avoids the explicit construction and inversion of the approximate operators ($\Lambda^*$) central to ALI. This feature is potentially advantageous for complex scenarios where formulating effective operators is challenging.

While promising, this work serves as a proof of concept and several natural extensions exist. Although the trained optimal SAC policy is highly efficient, the acceleration gains are only meaningful if the policy can be shown to generalize to different environments, without the need for expensive pre-training. Furthermore, the system should be submerged into a more realistic complex environment such as that provided by the Lightweaver code \cite{lightweaver_2021}. The effect the discount factor has on the policy as well as the density of the reward signal provided to the actor should be numerically tested. We hypothesise that reward sparse configurations, where the inner loop's reward is only dispensed to the agent at modular intervals in Fig. \ref{rl_loop_image}, would result in more optimal convergence at the price of increased compute but also decreased generalization. On the other hand, a reward-rich scenario could promote generalization, as the agent's policy is more tightly constrained and tethered to the behaviour of the underlying simulation. In the case where the system is out of equilibrium, a general imitation learning approach can be deployed \citep{imitation_leanring2024}.

Finally, we plan to extend the RL framework to related optimization problems in astrophysics that are easily "gamified", such as spectropolarimetric inversions with discrete node adjustments. It is our belief that reinforcement learning provides an exciting and flexible new paradigm for tackling complex, physics‐constrained problems within solar physics, enabling agents to learn directly from simulations and uncover efficient, non‑intuitive solutions.

\section{Acknowledgments}

The author thanks the University of Applied Sciences and Arts Northwestern Switzerland for providing the resources and support, particularly through conference opportunities, that greatly contributed to the development of this work. We would also like to thank Dr. Reza Kakooee for his expert knowledge of reinforcement learning and for early fruitful discussions. The machine learning aspects of the project were developed using PyTorch \citep{pytorch_2019} and the open-source Stable Baselines3 \citep{stable_baselines} reinforcement learning library, while the forward transfer model made use of a publicly available two-level Non-LTE code \citep{zmilic_2lvl_nlte_2023}.

\bibliography{references}{}

\begin{thebibliography}{38}
\expandafter\ifx\csname natexlab\endcsname\relax\def\natexlab#1{#1}\fi

\bibitem[Anusha et~al.(2009)Anusha, Nagendra, Paletou, \&
  Léger]{Bi_conjugate_Gradient_2009}
Anusha, L.~S., Nagendra, K.~N., Paletou, F., \& Léger, L., 2009.
\newblock Preconditioned bi-conjugate gradient method for radiative transfer in
  spherical media, {\it The Astrophysical Journal\/}, {\bf 704}(1), 661.

\bibitem[{Benedusi} et~al.(2023){Benedusi}, {Riva}, {Zulian},
  {{\v{S}}t{\v{e}}p{\'a}n}, {Belluzzi}, \& {Krause}]{matrix_free2023}
{Benedusi}, P., {Riva}, S., {Zulian}, P., {{\v{S}}t{\v{e}}p{\'a}n}, J.,
  {Belluzzi}, L., \& {Krause}, R., 2023.
\newblock {Scalable matrix-free solver for 3D transfer of polarized radiation
  in stellar atmospheres}, {\it Journal of Computational Physics\/}, {\bf 479},
  112013.

\bibitem[Berner et~al.(2019)Berner, Brockman, Chan, Cheung, Debiak, Dennison,
  Farhi, Fischer, Hashme, Hesse, J{\'o}zefowicz, Gray, Olsson, Pachocki,
  Petrov, de~Oliveira~Pinto, Raiman, Salimans, Schlatter, Schneider, Sidor,
  Sutskever, Tang, Wolski, \& Zhang]{Dota2019}
Berner, C., Brockman, G., Chan, B., Cheung, V., Debiak, P., Dennison, C.,
  Farhi, D., Fischer, Q., Hashme, S., Hesse, C., J{\'o}zefowicz, R., Gray, S.,
  Olsson, C., Pachocki, J.~W., Petrov, M., de~Oliveira~Pinto, H.~P., Raiman,
  J., Salimans, T., Schlatter, J., Schneider, J., Sidor, S., Sutskever, I.,
  Tang, J., Wolski, F., \& Zhang, S., 2019.
\newblock Dota 2 with large scale deep reinforcement learning, {\it ArXiv\/},
  {\bf abs/1912.06680}.

\bibitem[{Chappell} \& {Pereira}(2022)]{sunny_net2022}
{Chappell}, B.~A. \& {Pereira}, T. M.~D., 2022.
\newblock {SunnyNet: Neural network framework for solving 3D NLTE radiative
  transfer in stellar atmospheres}, Astrophysics Source Code Library, record
  ascl:2202.024.

\bibitem[Chen et~al.(2022)Chen, Jeffery, Zhong, McClenny, Braga-Neto, \&
  Wang]{supernova2022}
Chen, X., Jeffery, D.~J., Zhong, M., McClenny, L., Braga-Neto, U., \& Wang, L.,
  2022.
\newblock {Using Physics Informed Neural Networks for Supernova Radiative
  Transfer Simulation}.

\bibitem[{D{\'\i}az Baso} et~al.(2025){D{\'\i}az Baso}, {Asensio Ramos}, {de la
  Cruz Rodr{\'\i}guez}, {da Silva Santos}, \& {Rouppe van der
  Voort}]{carlos_neural_fields_inversions_2025}
{D{\'\i}az Baso}, C.~J., {Asensio Ramos}, A., {de la Cruz Rodr{\'\i}guez}, J.,
  {da Silva Santos}, J.~M., \& {Rouppe van der Voort}, L., 2025.
\newblock {Exploring spectropolarimetric inversions using neural fields: Solar
  chromospheric magnetic field under the weak-field approximation}, {\it
  \aap\/}, {\bf 693}, A170.

\bibitem[Fawzi et~al.(2022)Fawzi, Balog, Huang, Hubert, Romera-Paredes,
  Barekatain, Novikov, Ruiz, Schrittwieser, Swirszcz, Silver, Hassabis, \&
  Kohli]{AlphaTensor2022}
Fawzi, A., Balog, M., Huang, A., Hubert, T., Romera-Paredes, B., Barekatain,
  M., Novikov, A., Ruiz, F. J.~R., Schrittwieser, J., Swirszcz, G., Silver, D.,
  Hassabis, D., \& Kohli, P., 2022.
\newblock Discovering faster matrix multiplication algorithms with
  reinforcement learning, {\it Nature\/}, {\bf 610}(7930), 47--53.

\bibitem[{Gudiksen} et~al.(2011){Gudiksen}, {Carlsson}, {Hansteen}, {Hayek},
  {Leenaarts}, \& {Mart{\'\i}nez-Sykora}]{BIFROST}
{Gudiksen}, B.~V., {Carlsson}, M., {Hansteen}, V.~H., {Hayek}, W., {Leenaarts},
  J., \& {Mart{\'\i}nez-Sykora}, J., 2011.
\newblock {The stellar atmosphere simulation code Bifrost. Code description and
  validation}, {\it \aap\/}, {\bf 531}, A154.

\bibitem[Haarnoja et~al.(2018{\natexlab{a}})Haarnoja, Zhou, Abbeel, \&
  Levine]{sac_2018}
Haarnoja, T., Zhou, A., Abbeel, P., \& Levine, S., 2018{\natexlab{a}}.
\newblock Soft actor-critic: Off-policy maximum entropy deep reinforcement
  learning with a stochastic actor, in {\em International conference on machine
  learning\/}, pp. 1861--1870, PMLR.

\bibitem[Haarnoja et~al.(2018{\natexlab{b}})Haarnoja, Zhou, Hartikainen,
  Tucker, Ha, Tan, Kumar, Zhu, Gupta, Abbeel, et~al.]{sac_applications_2018}
Haarnoja, T., Zhou, A., Hartikainen, K., Tucker, G., Ha, S., Tan, J., Kumar,
  V., Zhu, H., Gupta, A., Abbeel, P., et~al., 2018{\natexlab{b}}.
\newblock Soft actor-critic algorithms and applications, {\it arXiv preprint
  arXiv:1812.05905\/}.

\bibitem[Hill et~al.(2018)Hill, Raffin, Ernestus, Gleave, Kanervisto, Traore,
  Dhariwal, Hesse, Klimov, Nichol, Plappert, Radford, Schulman, Sidor, \&
  Wu]{stable_baselines}
Hill, A., Raffin, A., Ernestus, M., Gleave, A., Kanervisto, A., Traore, R.,
  Dhariwal, P., Hesse, C., Klimov, O., Nichol, A., Plappert, M., Radford, A.,
  Schulman, J., Sidor, S., \& Wu, Y., 2018.
\newblock Stable baselines, \url{https://github.com/hill-a/stable-baselines}.

\bibitem[{Jarolim} et~al.(2025){Jarolim}, {Molnar}, {Tremblay}, {Centeno}, \&
  {Rempel}]{momo_pinns_2025}
{Jarolim}, R., {Molnar}, M.~E., {Tremblay}, B., {Centeno}, R., \& {Rempel}, M.,
  2025.
\newblock {PINN ME: A Physics-Informed Neural Network Framework for Accurate
  Milne-Eddington Inversions of Solar Magnetic Fields}, {\it arXiv e-prints\/},
  p. arXiv:2502.13924.

\bibitem[Jumper et~al.(2021)Jumper, Evans, Pritzel, Green, Figurnov,
  Ronneberger, Tunyasuvunakool, Bates, Žídek, Potapenko, Bridgland, Meyer,
  Kohl, Ballard, Cowie, Romera-Paredes, Nikolov, Jain, Adler, Back, Petersen,
  Reiman, Clancy, Zielinski, Steinegger, Pacholska, Berghammer, Bodenstein,
  Silver, Vinyals, Senior, Kavukcuoglu, Kohli, \& Hassabis]{alphafold2021}
Jumper, J., Evans, R., Pritzel, A., Green, T., Figurnov, M., Ronneberger, O.,
  Tunyasuvunakool, K., Bates, R., Žídek, A., Potapenko, A., Bridgland, A.,
  Meyer, C., Kohl, S. A.~A., Ballard, A.~J., Cowie, A., Romera-Paredes, B.,
  Nikolov, S., Jain, R., Adler, J., Back, T., Petersen, S., Reiman, D., Clancy,
  E., Zielinski, M., Steinegger, M., Pacholska, M., Berghammer, T., Bodenstein,
  S., Silver, D., Vinyals, O., Senior, A.~W., Kavukcuoglu, K., Kohli, P., \&
  Hassabis, D., 2021.
\newblock Highly accurate protein structure prediction with alphafold, {\it
  Nature\/}, {\bf 596}(7873), 583--589.

\bibitem[Korber et~al.(2023)Korber, Bianco, Tolley, \& Kneib]{Michele2023}
Korber, D., Bianco, M., Tolley, E., \& Kneib, J.-P., 2023.
\newblock {PINION: physics-informed neural network for accelerating radiative
  transfer simulations for cosmic reionization}, {\it Monthly Notices of the
  Royal Astronomical Society\/}, {\bf 521}(1), 902--915.

\bibitem[Lagerquist et~al.(2021)Lagerquist, Turner, Ebert-Uphoff, Stewart, \&
  Hagerty]{PINN2021_weather}
Lagerquist, R., Turner, D., Ebert-Uphoff, I., Stewart, J., \& Hagerty, V.,
  2021.
\newblock Using deep learning to emulate and accelerate a radiative transfer
  model, {\it Journal of Atmospheric and Oceanic Technology\/}, {\bf 38}(10),
  1673 -- 1696.

\bibitem[Lillicrap et~al.(2015)Lillicrap, Hunt, Pritzel, Heess, Erez, Tassa,
  Silver, \& Wierstra]{moving_target2015}
Lillicrap, T.~P., Hunt, J.~J., Pritzel, A., Heess, N., Erez, T., Tassa, Y.,
  Silver, D., \& Wierstra, D., 2015.
\newblock Continuous control with deep reinforcement learning, {\it arXiv
  preprint arXiv:1509.02971\/}.

\bibitem[{Mihalas}(1978)]{Mihalasbook1978}
{Mihalas}, D., 1978.
\newblock {\it {Stellar atmospheres}\/}.

\bibitem[Milic(2023)]{zmilic_2lvl_nlte_2023}
Milic, I., 2023.
\newblock 2lvl\_nlte: Pedagogical non-lte radiative transfer with a two-level
  atom, \url{https://github.com/ivanzmilic/2lvl_nlte}, Accessed: 2025-04-14.

\bibitem[{Mili{\'c}} \& {Atanackovi{\'c}}(2014)]{Implicit_Lambda2014}
{Mili{\'c}}, I. \& {Atanackovi{\'c}}, O., 2014.
\newblock {Accelerating NLTE radiative transfer by means of the Forth-and-Back
  Implicit Lambda Iteration: A two-level atom line formation in 2D Cartesian
  coordinates}, {\it Advances in Space Research\/}, {\bf 54}(7), 1297--1307.

\bibitem[Mnih et~al.(2015)Mnih, Kavukcuoglu, Silver, Rusu, Veness, Bellemare,
  Graves, Riedmiller, Fidjeland, Ostrovski, et~al.]{moving_target2015human}
Mnih, V., Kavukcuoglu, K., Silver, D., Rusu, A.~A., Veness, J., Bellemare,
  M.~G., Graves, A., Riedmiller, M., Fidjeland, A.~K., Ostrovski, G., et~al.,
  2015.
\newblock Human-level control through deep reinforcement learning, {\it
  nature\/}, {\bf 518}(7540), 529--533.

\bibitem[Mu et~al.(2023)Mu, Chen, Yuan, \& Qin]{torch_adaptor2023}
Mu, B., Chen, L., Yuan, S., \& Qin, B., 2023.
\newblock A radiative transfer deep learning model coupled into wrf with a
  generic fortran torch adaptor, {\it Frontiers in Earth Science\/}, {\bf 11}.

\bibitem[Novati et~al.(2021)Novati, de~Laroussilhe, \&
  Koumoutsakos]{turbulence2020}
Novati, G., de~Laroussilhe, H.~L., \& Koumoutsakos, P., 2021.
\newblock Automating turbulence modelling by multi-agent reinforcement
  learning, {\it Nature Machine Intelligence\/}, {\bf 3}(1), 87--96.

\bibitem[Olson \& Kunasz(1987)]{OLSON1987325}
Olson, G.~L. \& Kunasz, P., 1987.
\newblock Short characteristic solution of the non-lte line transfer problem by
  operator perturbation—i. the one-dimensional planar slab, {\it Journal of
  Quantitative Spectroscopy and Radiative Transfer\/}, {\bf 38}(5), 325--336.

\bibitem[Osborne \& Mili{\'c}(2021)]{lightweaver_2021}
Osborne, C.~M. \& Mili{\'c}, I., 2021.
\newblock The lightweaver framework for nonlocal thermal equilibrium radiative
  transfer in python, {\it The Astrophysical Journal\/}, {\bf 917}(1), 14.

\bibitem[Paszke et~al.(2019)Paszke, Gross, Massa, Lerer, Bradbury, Chanan,
  Killeen, Lin, Gimelshein, Antiga, Desmaison, Kopf, Yang, DeVito, Raison,
  Tejani, Chilamkurthy, Steiner, Fang, Bai, \& Chintala]{pytorch_2019}
Paszke, A., Gross, S., Massa, F., Lerer, A., Bradbury, J., Chanan, G., Killeen,
  T., Lin, Z., Gimelshein, N., Antiga, L., Desmaison, A., Kopf, A., Yang,
  E.~Z., DeVito, Z., Raison, M., Tejani, A., Chilamkurthy, S., Steiner, B.,
  Fang, L., Bai, J., \& Chintala, S., 2019.
\newblock Pytorch: An imperative style, high-performance deep learning library,
  in {\em Advances in Neural Information Processing Systems 32\/}, vol.~32, pp.
  8024--8035, Curran Associates, Inc.

\bibitem[Przybylski et~al.(2022)Przybylski, Cameron, Solanki, Rempel,
  Leenaarts, Anusha, Witzke, \& Shapiro]{MuramChromo}
Przybylski, D., Cameron, R., Solanki, S., Rempel, M., Leenaarts, J., Anusha,
  L., Witzke, V., \& Shapiro, A., 2022.
\newblock Chromospheric extension of the muram code, {\it Astronomy \&
  Astrophysics\/}, {\bf 664}, A91.

\bibitem[{Rybicki} \& {Hummer}(1991)]{ali1}
{Rybicki}, G.~B. \& {Hummer}, D.~G., 1991.
\newblock {An accelerated lambda iteration method for multilevel radiative
  transfer. I. Non-overlapping lines with background continuum}, {\it \aap\/},
  {\bf 245}, 171--181.

\bibitem[{Rybicki} \& {Hummer}(1992)]{ali2}
{Rybicki}, G.~B. \& {Hummer}, D.~G., 1992.
\newblock {An accelerated lambda iteration method for multilevel radiative
  transfer. II. Overlapping transitions with full continuum.}, {\it \aap\/},
  {\bf 262}, 209--215.

\bibitem[{Rybicki} \& {Hummer}(1994)]{ali3}
{Rybicki}, G.~B. \& {Hummer}, D.~G., 1994.
\newblock {An accelerated lambda iteration method for multilevel radiative
  transfer. III. Noncoherent electron scattering}, {\it \aap\/}, {\bf 290},
  553--562.

\bibitem[{Sethuram} et~al.(2023){Sethuram}, {Cochrane}, {Hayward}, {Acquaviva},
  {Villaescusa-Navarro}, {Popping}, \& {Wise}]{rt_galaxy_2023}
{Sethuram}, S.~S., {Cochrane}, R.~K., {Hayward}, C.~C., {Acquaviva}, V.,
  {Villaescusa-Navarro}, F., {Popping}, G., \& {Wise}, J.~H., 2023.
\newblock {Emulating radiative transfer with artificial neural networks}, {\it
  \mnras\/}, {\bf 526}(3), 4520--4528.

\bibitem[{Silver} et~al.(2016){Silver}, {Huang}, {Maddison}, {Guez}, {Sifre},
  {van den Driessche}, {Schrittwieser}, {Antonoglou}, {Panneershelvam},
  {Lanctot}, {Dieleman}, {Grewe}, {Nham}, {Kalchbrenner}, {Sutskever},
  {Lillicrap}, {Leach}, {Kavukcuoglu}, {Graepel}, \& {Hassabis}]{GO2016}
{Silver}, D., {Huang}, A., {Maddison}, C.~J., {Guez}, A., {Sifre}, L., {van den
  Driessche}, G., {Schrittwieser}, J., {Antonoglou}, I., {Panneershelvam}, V.,
  {Lanctot}, M., {Dieleman}, S., {Grewe}, D., {Nham}, J., {Kalchbrenner}, N.,
  {Sutskever}, I., {Lillicrap}, T., {Leach}, M., {Kavukcuoglu}, K., {Graepel},
  T., \& {Hassabis}, D., 2016.
\newblock {Mastering the game of Go with deep neural networks and tree search},
  {\it \nat\/}, {\bf 529}(7587), 484--489.

\bibitem[Silver et~al.(2017)Silver, Schrittwieser, Simonyan, Antonoglou, Huang,
  Guez, Hubert, Baker, Lai, Bolton, Chen, Lillicrap, Hui, Sifre, van~den
  Driessche, Graepel, \& Hassabis]{AlphaZero2017}
Silver, D., Schrittwieser, J., Simonyan, K., Antonoglou, I., Huang, A., Guez,
  A., Hubert, T., Baker, L., Lai, M., Bolton, A., Chen, Y., Lillicrap, T., Hui,
  F., Sifre, L., van~den Driessche, G., Graepel, T., \& Hassabis, D., 2017.
\newblock Mastering the game of go without human knowledge, {\it Nature\/},
  {\bf 550}(7676), 354--359.

\bibitem[{Steiner}(1991)]{Grid1_1991}
{Steiner}, O., 1991.
\newblock {Fast solution of radiative transfer problems using a method of
  multiple grids}, {\it \aap\/}, {\bf 242}(1), 290--300.

\bibitem[{Trujillo Bueno} \& {Fabiani Bendicho}(1995)]{gs_method1995}
{Trujillo Bueno}, J. \& {Fabiani Bendicho}, P., 1995.
\newblock {A Novel Iterative Scheme for the Very Fast and Accurate Solution of
  Non-LTE Radiative Transfer Problems}, {\it \apj\/}, {\bf 455}, 646.

\bibitem[Uitenbroek(2001)]{RH2001}
Uitenbroek, H., 2001.
\newblock Multilevel radiative transfer with partial frequency redistribution,
  {\it The Astrophysical Journal\/}, {\bf 557}(1), 389.

\bibitem[{Vicente Ar{\'e}valo} et~al.(2022){Vicente Ar{\'e}valo}, {Asensio
  Ramos}, \& {Esteban Pozuelo}]{Graph2022}
{Vicente Ar{\'e}valo}, A., {Asensio Ramos}, A., \& {Esteban Pozuelo}, S., 2022.
\newblock {Accelerating Non-LTE Synthesis and Inversions with Graph Networks},
  {\it \apj\/}, {\bf 928}(2), 101.

\bibitem[{{\v{S}}t{\v{e}}p{\'a}n} \& {Trujillo Bueno}(2013)]{Grid2_2013}
{{\v{S}}t{\v{e}}p{\'a}n}, J. \& {Trujillo Bueno}, J., 2013.
\newblock {PORTA: A three-dimensional multilevel radiative transfer code for
  modeling the intensity and polarization of spectral lines with massively
  parallel computers}, {\it \aap\/}, {\bf 557}, A143.

\bibitem[Zare et~al.(2024)Zare, Kebria, Khosravi, \&
  Nahavandi]{imitation_leanring2024}
Zare, M., Kebria, P.~M., Khosravi, A., \& Nahavandi, S., 2024.
\newblock A survey of imitation learning: Algorithms, recent developments, and
  challenges, {\it IEEE Transactions on Cybernetics\/}.

\end{thebibliography}
\bibliographystyle{rasti}

\label{lastpage}
\end{document}